\documentclass[flushrt,preprint]{aastex}
\usepackage{amssymb,amsmath,graphicx,psfrag}
\usepackage{url}
\begin{document}
\title{How Soft Gamma Repeaters {Might} Make Fast Radio Bursts}
\shorttitle{How SGR Make FRB}
\shortauthors{Katz}
\author{J. I. Katz\altaffilmark{}}
\affil{Department of Physics and McDonnell Center for the Space Sciences}
\affil{Washington University, St. Louis, Mo. 63130 USA}
\email{katz@wuphys.wustl.edu}
\begin{abstract}
There are several phenomenological similarities between Soft Gamma Repeaters
and Fast Radio Bursts, including duty factors, time scales and repetition.
The sudden release of magnetic energy in a neutron star magnetosphere, as in
popular models of SGR, can meet the energy requirements of FRB but requires
both the presence of magnetospheric plasma in order that dissipation occur
in a transparent region and a mechanism for releasing much of that energy
quickly.  FRB sources and SGR are distinguished by long-lived (up to
thousands of years) current-carrying coronal arches remaining from formation
of the young neutron star, and their decay ends the phase of SGR/AXP/FRB
activity even though ``magnetar'' fields may persist.  Runaway increase in
resistance when the current density exceeds a threshold releases
magnetostatic energy in a sudden burst and produces high brightness GHz
emission of FRB by a coherent process; SGR are produced when released energy
thermalizes as an equlibrium pair plasma.  Failures of some alternative FRB
models and the non-detection of SGR~1806-20 at radio frequencies are
discussed in appendices.
\end{abstract}
\maketitle
\section{Introduction}
Fast Radio Bursts (FRB) and Soft Gamma Repeaters (SGR) are rare and brief
episodic events.  Many authors \citep{PP07,PP13,Ku14,L14,Ku15,PC15,K16} have
noted this similarity and have suggested that FRB and SGR may be associated.
This paper considers the physical processes that may make FRB from the
sudden releases of magnetic energy that are believed to power SGR.  FRB are,
even at ``cosmological'' distances, $\sim 10^{-4}$ as powerful as the most
powerful SGR, so a process that makes FRB from SGR need not be efficient.

The recent discovery \citep{M15} of linear polarization and Faraday rotation
in one FRB strengthens the case for cosmological distances.  The measured
rotation measure of FRB 110523 was $-186$ rad/m$^2$, implying the
line-of-sight integral
\begin{equation}
\int n_e B_\parallel\,d\ell = 229\ \text{pc $\mu$G cm}^{-3}.
\end{equation}
Comparing to its dispersion measure of 623 pc cm$^{-3}$ yields the electron
density-averaged parallel field along the line-of-sight
\begin{equation}
\langle B_\parallel \rangle_{n_e} \equiv {\int n_e B_\parallel\,d\ell \over
\int n_e\,d\ell} = 0.37\ \mu\mathrm{G}.
\end{equation}
This is more than an order of magnitude less than typical spiral galaxy
fields $\sim 10\,\mu$G \citep{W02}, and several orders of magnitude less
than plausible fields in dense clouds or the immediate environments of
stars.  It indicates that the overwhelming majority of the dispersion occurs
in the intergalactic medium where nanogauss or weaker fields are expected,
and confirms the inference of ``cosmological'' distances.  {This implies
that FRB may have powers as great as $10^{43}$ erg/s \citep{T13b}, and
requires correspondingly energetic sources.  SGR, observed to have powers as
high as $10^{47}$ erg/s \citep{H05}, can satisfy this requirement.}

It is generally accepted \citep{K82,TD92,TD95,K96} that SGR outbursts result
from the dissipation of magnetostatic energy in the magnetosphere of a
highly magnetized neutron star (the ``magnetar'' model).  \cite{TD95}
suggested that the rare giant flares of SGR are produced by a rupture
propagating across the entire solid crust, the general failure of a brittle
object found in a model of brittle fracture \citep{K86,BTW87} now called
``self organized criticality''.
{\S2 presents the phenomenological case for associating SGR with FRB.}
In \S3 I apply the well-known theory of long-lived coronal {currents}
in SGR to FRB.  These {currents} are relics of neutron star formation,
and their lifetimes are consistent with the inferred ages of SGR and
indicate that FRB are similarly young.
The presence of such magnetospheric currents
distinguishes these magnetars from neutron stars in which currents are
confined to the dense interior.  \S4 estimates the energy to which electrons
are accelerated, which determines the decay time of the magnetospheric
currents.  \S5 suggests a possible outburst mechanism and \S6 considers
curvature radiation as the emission mechanism of FRB, estimating the
required charge clumping factor.  \S7 is a brief summary and conclusion.
Appendix A considers the hypothesis that Compton recoil of pair annihilation
radiation may produce a plasma instability in the irradiated plasma, leading
to large amplitude plasma waves that might produce high brightness GHz
radiation; this hypothesis fails on energetic grounds.
Appendix B considers some alternatives to the SGR hypothesis for FRB at
cosmological distances, specifically giant pulsar pulses and neutron star
collapse, and finds them wanting.  Appendix C discusses the non-detection
of a FRB in a radio observation \citep{TKP16} fortuitously simultaneous with
the giant outburst of SGR~1806-20, and suggests possible explanations.
\section{The Case for Associating SGR and FRB}
FRB and SGR have three distinct similarities that suggest an association:

1. {Duty Factor}: Their duty factors, defined $D \equiv \langle F(t)
\rangle^2/\langle F(t)^2 \rangle$, where $F(t)$ is the flux, quantify the
the fraction of the time in which a source emits at close to its peak flux,
are extremely low: $D \sim 10^{-10}$ for SGR and $D < 10^{-8}$ for at least
one FRB \citep{L15}.

2. {Time scale}: The intrinsic durations of FRB have not been measured but
\cite{T13b} found instrumentally-limited upper bounds of about 1 ms for
several FRB.  Some other FRB have had widths up to $\approx 10\,$ms that are
attributed, because of their $\propto \nu^{-4}$ frequency dependence, to
broadening by multipath propagation, and only upper bounds can be placed on
the intrinsic pulse widths.  This is consistent with the rise times of giant
SGR outbursts: The rise time of the March 5, 1979 outburst of SGR~{0525-66}
was $< 200\,\mu$s \citep{C80a,C80b}, \cite{P05} reported an exponential
rise time of 300$\,\mu$s for the giant December~27, 2004 outburst of
SGR~1806-20, while their published data suggest a value of 200$\,\mu$s, and
the giant August~27, 1998 flare of SGR~1900+14 had a rise time of $< 4\,$ms
\citep{H99} and earlier outbursts had rise times $\le 8\,$ms \citep{M79}.
These time scales are shorter than those of any other known astronomical
event except gravitational wave emission by coalescing black holes and the
pulses and subpulses of some pulsars; GRB durations and subpulse time scales
are all $\gtrsim 30\,$ms \citep{F94,Q13}.

3. {Repetition}: SGR repeat in complex irregular patterns, with periods of
activity interspersed in longer periods of quiescence.  The double pulse of
FRB 121002, with subpulses separated by about 2 ms \citep{T13a,C15} may be
considered a repetition, and multiple repetitions of FRB 121102 were
recently discovered \citep{S16}, with irregular spacings reminiscent of the
activity of SGR 1806-20 \citep{L87}.  FRB are not catastrophic events that
destroy their sources, and resemble SGR rather than GRB.

\section{The Magnetosphere}
\label{magnetosphere}
It was realized soon after the discovery of the first SGR~{0525-66} in 1979
that its combination of rapid rise ($< 200\,\mu$s) and energy release of
$10^{44}$--$10^{45}$ ergs \citep{C80a,C80b} required a source in a region of
high energy density.  Magnetic reconnection \citep{PF00} in the
magnetosphere of a neutron star with (then) unprecedentedly strong magnetic
fields was a natural model for the energy source \citep{K82,TD92,TD95}.

In magnetic reconnection magnetic energy is dissipated by resistive heating
or particle acceleration in thin current sheets separating regions of
differing, but comparatively homogeneous, magnetic fields.  The current
sheets may be modeled as resistors in series with an inductive energy store.
If the electric field in the current sheets is large enough,
counterstreaming electrons and ions (or positrons if they are present) make
the plasma unstable to a variety of plasma waves.  Correlated particles
(``clumps'') become effective scatterers, a condition described by an
``anomalous'' resistivity.  This anomalous resistivity may be many orders of
magnitude greater than the nominal resistivity resulting from interactions
with uncorrelated particles.

The explosive growth of the resistivity $\rho_e$ as a plasma instability
exponentiates, while the current is held nearly constant by the circuit
inductance, implies a proportionally explosive growth of the electric field
${\vec E}=\rho_e {\vec J}$ and of the dissipation rate ${\vec E} \cdot
{\vec J} = \rho_e J^2$ (in large magnetic fields $\rho_e$, written here as a
scalar, is a tensor, complicating the problem but not changing the
qualitative conclusion).  The result can be rapid dissipation of large
{amounts} of magnetostatic energy, depending on the evolution of
$\rho_e$ in space and time.


At energy fluxes $\gtrsim 10^{29}$ ergs/cm$^2$-s, such as observed in giant
SGR flares, the energy released thermalizes into an opaque equilibrium
pair-black body photon gas, and a black body spectrum with temperature
$\sim 20$ keV is emitted \citep{K96}, roughly consistent with observed SGR
spectra.  Confinement of the charged particles by the magnetic field
\citep{K82,TD95,K96} permits the radiated intensity to exceed the Eddington
limit by large factors, as observed.  Equilibration is expected during most
of the $\sim 0.1\,$s duration of giant SGR flares.  Their initial $< 1\,$ms
rise times may correspond to the progress of a reconnection wave along a
current sheet as plasma instability produces large amplitude charge clumps
and increases resistivity by orders of magnitude.

Some positrons escape their source and annihilate in dense
cooler matter such as the neutron star's surface, emitting characteristic
annihilation radiation at energies about 511 keV, as reported from
SGR0525-66 \citep{C80a,C80b}.  It is natural to associate the $< 1\,$ms FRB
with the $< 1\,$ms rise time of giant SGR flares, before the
electron-positron pairs have had time to equilibrate as a lower temperature
plasma, converting most of their rest mass energy into black body photons.

{Idealized} neutron stars have sharply defined radii, with
thermal scale heights $\lesssim 1\,$cm after rapid early cooling.
Above this scale height (unless they are accreting) the surrounding space is 
nearly vacuum, filled only with a non-neutral plasma with the
Goldreich-Julian density \citep{D47,HB65,GJ69}
\begin{equation}
\label{GJ}
n_{GJ} = - {{\vec \omega_{rot}} \cdot {\vec B} \over 2 \pi c e}
\end{equation}
for an aligned magnetic dipole field and rotation rate $\omega_{rot}$.  The
defining characteristic of ``magnetar'' models is the assumption that
rotation is unimportant: their radiation is derived from their magnetostatic
energy, $n_{GJ}$ is insignificant and (except for rotational modulation
of the observed radiation) the star may be considered non-rotating.

This leads to questions for magnetar models of {FRB}:
Magnetic reconnection is implausible in the
neutron star interior whose high density implies high conductivity and low
electron-ion drift velocity.  This precludes plasma instability and the 
development of anomalous (turbulent) resistivity, the generally accepted
mechanism of magnetic reconnection \citep{PF00} in low density plasma.
  If reconnection were somehow to occur in the neutron star interior it
would only warm the dense matter there, with the released energy slowly
diffusing to the surface to be radiated as thermal X-rays.

The electron-ion drift speed
\begin{equation}
v_{drift} = {J \over n_e e} \sim {B c \over 4 \pi r n_e e} \sim 10^4
{B_{15} \over \rho_m}\ \text{cm/s},
\end{equation}
where $\rho_m$ is the mass density and $B_{15} \equiv B/(10^{15}
\mathrm{G})$.
Anomalous resistivity can only occur where $\rho_m \ll 1\,$g/cm$^3$, in an
optically thin atmosphere (that may not exist if magnetic quantization of
electron states gives neutron stars abrupt surfaces).

Rapidly rising SGR outbursts {and FRB} must occur in nearly transparent
regions (optical depth $\lesssim 1$) in order that their radiation escape in
their observed sub-ms rise times.  Yet reconnection cannot occur in a vacuum
magnetosphere, despite its high magnetostatic energy density, because no
currents flow in vacuum.  {Here the theory \citep{TD92,TD95,TLK02,BT07}
of the magnetar origin of SGR is applied to SGR models of FRB.}

These problems may be resolved if substantial portions of the neutron star's
magnetic moment are produced by current loops flowing through long-lived
quasi-neutral coronal arches above the high density neutron star surface.
Actual force-free configurations {(in magnetar fields the cross-field
conductivity is very small, electrons are in their quantized ground states
and any cross-field currents would imply enormous Lorentz forces, so that
${\vec J} \parallel {\vec B}$}) are complex \citep{P79,AMPCD16}.
%

As a neutron star forms from the collapse of a stellar core, frozen-in
magnetic fields are amplified by compression and possibly by turbulence.  In
the earlier stages of contraction the matter pressure far exceeds the
magnetic stress.  Currents flow along the field lines, along which the
conductivity is always higher than perpendicular to them (although at high
densities there may also be cross-field currents).  As the matter, radiating
neutrinos, gradually settles into its final configuration, the magnetic
stress becomes dominant in the low density regions above the developing
neutron star surface, and the matter there is constrained to flow along the
field lines.  Magnetic field lines still penetrate that surface (if they
didn't, all multipole moments would be zero).  The magnetosphere must be
force-free, with ${\vec J} \parallel {\vec B}$, because pressure gradients
are insufficient to oppose any Lorentz force and because the cross-field
conductivity is low at low matter density.  However, there is no reason to
expect the parallel component of $\vec J$ to disappear during collapse.

The resulting picture is one of a magnetosphere in which, near the stellar
surface
\begin{equation}
\label{currentdensity}
J \sim {cB \over 4 \pi r} \sim 2 \times 10^{18} B_{15}\ {\mathrm{esu} \over
\mathrm{cm^2 s}},
\end{equation}
where $B_{15}$ refers only to that portion of the field generated by
magnetospheric currents.

Current flows on loops, and produces a magnetic flux through a
representative loop
\begin{equation}
\label{flux}
\Phi \sim B r^2 \sim 10^{27} B_{15}\ \text{G-cm}^2.
\end{equation}
Magnetospheric current loops of this sort have been considered by
\cite{TLK02,BT07,B09} as the origin of magnetar coron\ae\ and by
\cite{L06,L13} to explain SGR flares and magnetar ``anti-glitches'' if they
open to infinity during flares, in analogy to Solar coronal mass ejections.

The electromotive force (EMF) induced by the changing flux through the loop
drives the current flow along the magnetic field lines 
At one foot of the arch electrons are pulled out of the surface plasma
and accelerated towards the other foot; positive ions (and possibly
positrons) are pulled from the surface and accelerated in the opposite
direction.  Both signs of charge carriers contribute to the current.
\cite{BT07} assumed only one positive charge carrier (or positive charge
carriers all with the same charge to mass ratio) and found that no steady
solution is possible and that pairs must be produced, but multiple positive
ion species and ionization states are likely to be present.  Our estimates
average over any rapid variability, whose time scale is $\sim r/c \sim
30\,\mu$s, shorter even than SGR rise times and the upper bounds on FRB
duration.

The current is mostly carried by relativistic electrons with density
\begin{equation}
\label{edensity}
n_e \sim {J \over (1 + J_+/J_-)ce} \sim {B \over 4 \pi r e} \sim 2 \times
10^{17} B_{15}\ \mathrm{cm}^{-3},
\end{equation}
where $J_+$ and $J_-$ are the current densities of positive and electron
charge carriers, respectively; $J_+/J_- \le 1$ because ions move more slowly
than electrons.  The electrons in the current loop must be quasi-neutralized
by ions or positrons because otherwise their electrostatic field $E \sim n_e
e r^3/r^2 \sim B/4\pi$ would be impossibly large.

This electron density may be compared to the co-rotation density
(Eq.~\ref{GJ}): $n_{GJ}/n_e < 2 \omega_{rot} r/c$.  The co-rotating charge
density is insignificant except near the co-rotation radius, where the
magnetic energy density and available power are very small.

The quasi-neutral low density current-carrying plasma is gravitationally
attracted to the star, and would slide down the magnetic field lines to the
surface.  However, $n_e$ cannot fall below the value given by
Eq.~\ref{edensity} because that would interrupt the flow of current,
changing the flux through the loop and inducing an EMF sufficient to
maintain the current.

The current-carrying electrons strike the star with a Lorentz factor
$\gamma$, acquired in their descent from the top of the loop by the electric
field that lifts positive ions to maintain quasi-neutrality.  The implied
EMF is $\approx \gamma m_e c^2/e$ because contributions from the electric
field lifting the electrons and inside the star are small (electrons require
little gravitational energy to lift and the stellar interior is an Ohmic
conductor of high conductivity).  Equating this EMF to $\partial\Phi/
\partial(ct) \sim \Phi/c \tau \sim B r^2/c\tau$ yields the magnetic decay
time $\tau$:
\begin{equation}
\label{Bdecay}
\tau \sim {r^2 e B \over \gamma m_e c^3} \sim 6 \times 10^5 {B_{15} \over
\gamma}\ \mathrm{y}.
\end{equation}
Of course, only the component of field attributable to magnetospheric
current decays on this time scale.

\begin{figure}
\centering
\includegraphics[width=3.5in]{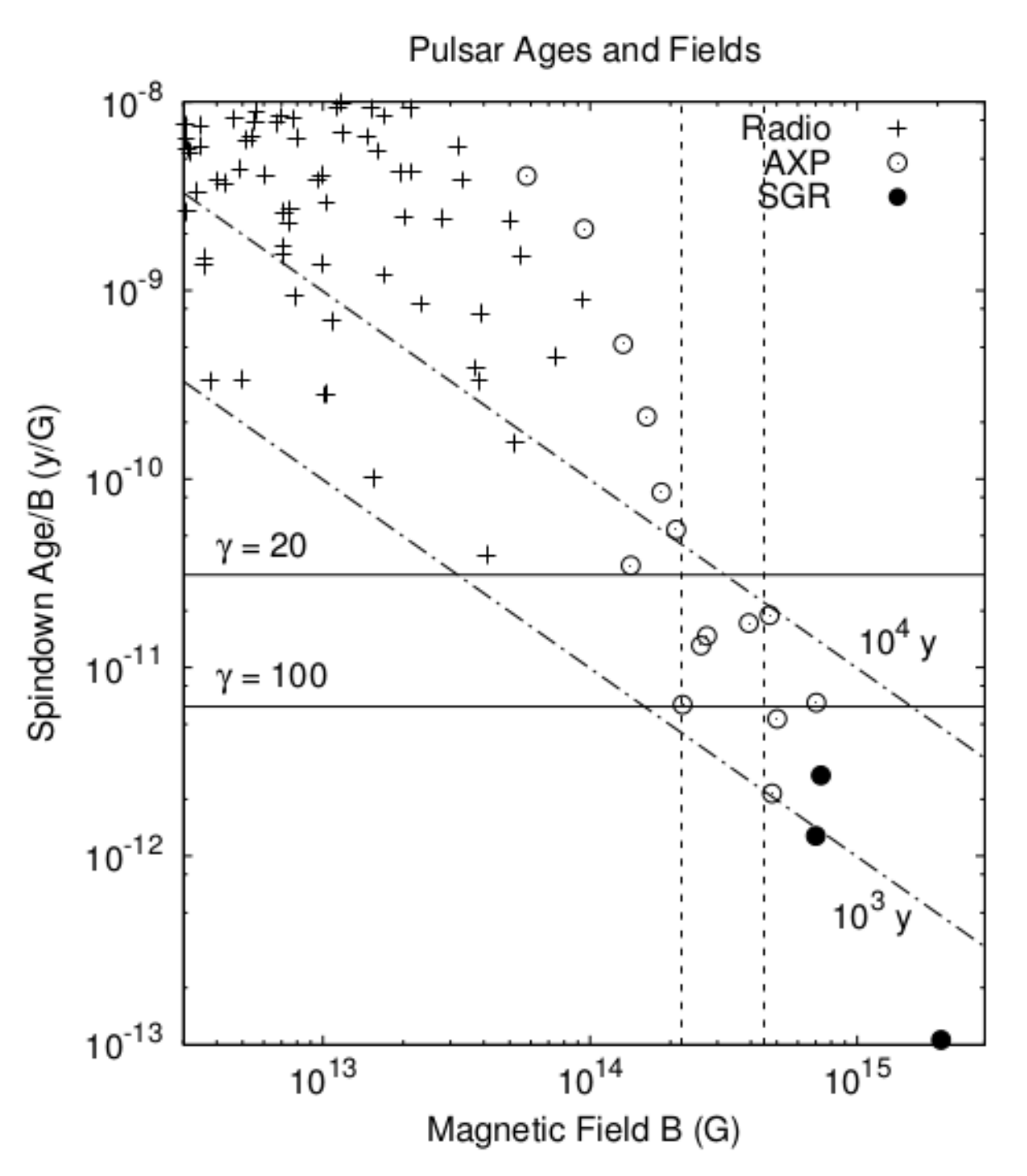}
\caption{\label{psrageB}Spindown ages and magnetic fields of all objects in
the ATNF Pulsar Catalogue \citep{MHTH05} with parameters in the displayed
range that are categorized as ``radio but not AXP'' or ``AXP''.  Of the 
``AXP'', only four (all with $2 \times 10^{14} < B < 3 \times 10^{14}$) have
reported radio emission.  The SGR that have shown giant flares (SGR~0526-66,
SGR~1806-20 and SGR~1900+14) are categorized as ``AXP'' but here shown with
a distinctive symbol.  Diagonal dot-dashed lines indicate spindown ages.
Horizontal lines indicate the ratio of lifetime to $B$ (plotted on the
``Spindown Age/B'' axis) corresponding to their value of $\gamma$.  All
``AXP'' lying below the $\gamma = 20$ line have $B > 2.2 \times 10^{14}$ G
and all below the $\gamma = 100$ line have $B > 4.5 \times 10^{14}$ G; these
fields are indicated by vertical dashed lines.}
\end{figure}

The ratio $\text{Age}/B$, where Age is the spin-down age, is plotted in
Fig.~\ref{psrageB} for pulsars from the ATNF Pulsar Catalogue
\citep{MHTH05}.  This provides an estimate of $\gamma$ for SGR/AXP for which
the magnetospheric current contributes a substantial fraction of the
slowing-down torque, but it has no such significance for ordinary radio
pulsars whose magnetospheric currents have decayed.  Pulsars categorized in
the Catalogue as ``radio but not AXP'' are indicated as ``Radio'' while
those categorized as ``AXP'' are indicated as ``AXP'' or (the three with
observed giant flares) ``SGR''.  Pulsars with smaller
estimated fields (including most ordinary pulsars as well as millisecond
pulsars) are not shown.  It is likely that each category contains some
incorrectly classified objects ({\it e.g.,\/} accreting neutron stars or
other X-ray pulsars misclassified as AXP in the Catalogue).

The horizontal lines are upper bounds on the actual ages corresponding to
the indicated values of $\gamma$.  If the actual ages are close to the
spindown ages, as is the case for pulsars that have spun down from much
faster rotation at birth, then SGR/FRB will lie on or below the lines
corresponding to their values of $\gamma$.  They will be below these
horizontal lines if their magnetospheric currents have decayed very little,
possibly suggesting that the neutron star is in the early stages of its
active life as a SGR/FRB.  If its spindown age is greater than its actual
age, a neutron star may lie above the horizontal line; however, the absence
of points in the upper right part of the figure suggests that this is not
common.

All radio pulsars lie above the $\gamma = 20$ line, suggesting that when
SGR reach the age $\tau$ they turn into radio pulsars even if their fields
remain in the ``magnetar'' range.  This is consistent with the hypothesis
that SGR must be younger than $\tau$, and that any older object can only be
a radio pulsar, however large its magnetic field.  Every ``AXP'' in the
Catalogue above the $\gamma = 20$ line has $B \le 2.2 \times 10^{14}$ G
(indicated by a dashed line), and every ``AXP'' with $B \le 2.2 \times
10^{14}$ is above the $\gamma = 20$ line.  This 1:1 correspondence suggests
that these objects form a population distinct from those of the higher-B,
younger-aged ``AXP'', and may not be SGR/AXP at all.

The concentration of the SGR with giant flares to the lower right corner of
Fig.~\ref{psrageB} appears to be statistically significant.  Of the five
``AXP'' below the $\gamma = 100$ line, three are the SGR with giant flares
($P = 0.024$) and these three are among the four ``AXP'' with the highest
Age/$B$ ($P = 0.011$).  Although these statistics are {\it a posteriori\/},
they suggest that as SGR age past $\tau$ they may remain AXP as a result of
magnetic dissipation within their interiors, but that their SGR activity 
decays along with their magnetospheric currents.  Only a fraction of AXP are
SGR, and the absence of SGR activity in most AXP (those with greater ratios
of age to field) is not an artefact of limited observations but an intrinsic
property. 

\section{Electron Energy}
\label{eenergy}
For a magnetic arch with a gravitational potential difference $\Omega$
between its top and the neutron star surface,
\begin{equation}
\label{EMF}
\int_A^C {\vec E} \cdot {\vec {d\ell}}_{arch} = \Omega {Am_p \over Ze}
\sim {G M m_p \delta r \over r^2 e} {A \over Z}
\end{equation}
is required to raise the neutralizing ions from A to the top of the arch (C)
a height $\delta r$ above the neutron star's surface,
where the atomic number $A$, proton mass $m_p$ and ionization state $Z$
describe the neutralizing ions.  The electrons, closing their path to the
stellar surface, acquire a kinetic energy $-e\int_C^A {\vec E} \cdot {\vec
{d\ell}}_{arch}$ that they dissipate in collisions in the star at A.

Above the surface of a cooling neutron star with a nominal surface black
body temperature of $\sim$ 0.1--1 keV iron may be $\sim$ 20--24 times
ionized, although less in strong magnetic fields that increase the
ionization potentials.  The plausible range of $A/Z$ is from 1 (accreted
hydrogen) to $\sim$ 2--5 for strongly magnetized iron.

The magnetostatic energy available (in a dipole field) is $B^2 R^3/6$
\citep{K82} and the dissipated power is $B^2 R^3/(6 \tau) \sim 10^{34}
B_{15}\gamma$ ergs/s if the neutron star is young enough that much of its
magnetic moment is produced by currents in loops that extend through the
surface into the magnetosphere.  After those loops decay, the stars will
retain the magnetic moments produced by their internal currents.  The same
result is obtained, to order of magnitude, from Eq.~\ref{currentdensity} if
$J$ is assumed to extend over the entire surface:
\begin{equation}
\label{power}
P \sim 4 \pi r^2 n_e \gamma m_e c^3 \sim {B r \gamma m_e c^3 \over e} \sim
6 \times 10^{34} \gamma B_{15}\ \text{ergs/s},
\end{equation}
consistent with the steady (except for rotational modulation) X-ray emission
from the heated surface.  This power is in addition to that generated by
magnetic dissipation in the stellar interior.

Equating the EMF (Eq.~\ref{EMF}) to $\gamma m_e c^2/e$ and taking $\delta r =
3\,$km,
\begin{equation}
\label{gammaeq}
\gamma \approx {\Omega \over c^2} {m_p \over m_e} {A \over Z} \approx 40
{A \over Z},
\end{equation}
with $\gamma$ plausibly in the range 40--400.  Substituting this into
Eq.~\ref{Bdecay} yields
\begin{equation}
\label{tau}
\tau \sim {r^2 e B \over m_p c \Omega} {Z \over A} \sim 5000\ B_{15}
\left({\delta r \over 3\,\mathrm{km}}\right)^{-1} {Z \over A}\ \mathrm{y}.
\end{equation}


\section{Outburst Mechanism}
\subsection{Plasma Instability?}
The counterstreaming electron and ion currents would seem to invite the
two-stream plasma instability \citep{C74}.  This is unlikely to explain the
outbursts:
\begin{enumerate}
\item The plasma frequency corresponding to the electron density of
Eq.~\ref{edensity} exceeds observed FRB frequencies by orders of
magnitude, even allowing for the relativistic increase of mass with the
Lorentz factor (\ref{gammaeq}), so that oscillations of this plasma cannot
emit the observed radiation.
\item The counter streaming currents are present at all times.  An
instability might produce steady emission, but there is no evident mechanism
for it to make rare, low duty factor, giant outbursts.
\item The electrons are highly relativistic, moving with Lorentz factors
$\gamma \gg 1$ exactly parallel to $\vec B$ because their transverse motion
decays immediately in a quantizing magnetic field.  Their response to the
electric field of a longitudinal plasma wave with ${\vec E} \parallel
{\vec B}$ is reduced by a factor $\gamma^{-3}$ compared to that of
nonrelativistic electrons (their response to any transverse component of
$\vec E$ is zero because the magnetic field is quantizing).  This suppresses
the growth of the two-stream instability \citep{MY16}.
\end{enumerate}
\subsection{Coulomb Scattering}
The scattering cross-section of the neutralizing ions for the
current-carrying electrons may be estimated, though a quantitative
calculation would require use of their wave-functions in the quantizing
magnetic field.  We assume that the electron motion is only parallel to the
field and that the backscattering cross section is the integral of the Mott
cross-section in the relativistic limit, ignoring ionic recoil (ionic
transverse motion is less strongly quantized even in ``magnetar'' fields),
over the backward hemisphere:
\begin{align}
\begin{split}
\sigma &= 2 \pi \int_{-1}^0 {Z^2 e^4 \over 4 \gamma^2 m_e^2 c^4}
\csc^4{\!(\theta/2)} \cos^2{\!(\theta/2)}\,d\cos{\theta} \\ &= \pi (1 - \ln{2})
{Z^2 e^4 \over \gamma^2 m_e^2 c^4} = 7.6 \times 10^{-26}
{Z^2 \over \gamma^2}\ \mathrm{cm}^2.
\end{split}
\end{align}
Using the estimate (\ref{gammaeq}) for $\gamma$
\begin{equation}
\sigma \sim 5 \times 10^{-29} {Z^4 \over A^2}\ \mathrm{cm}^2 \lesssim 7
\times 10^{-27}\ \mathrm{cm}^2,
\end{equation}
where the upper bound is taken for completely ionized Fe ions.

The integrated column density of electrons along a coronal arch of length
$r$ is
\begin{equation}
n_e r \sim {B \over 4 \pi e} \sim 2 \times 10^{23} B_{15}\ \mathrm{cm}^{-2}.
\end{equation}
Assuming the current-carrying electrons are quasi-neutralized by ions (see
below), the corresponding column density of ions is 
\begin{equation}
\label{column}
n_i r ={n_e r \over Z} \sim {B \over 4 \pi e Z} \sim 2 \times 10^{23}
{B_{15} \over Z}\ \mathrm{cm}^{-2}.
\end{equation}

Comparing to the cross-section (\ref{column}) shows that the ionic Coulomb
scattering probability $P_{scatt}$ of the current-carrying electrons in the
magnetospheric arch is negligible:
\begin{align}
\begin{split}
\label{scatprob}
P_{scatt} &=
n_i r \sigma \sim {Z^2 e^4 \over \gamma^2 m_e^2 c^4} {B \over 4 \pi e Z}
\sim {1 - \ln{2} \over 4} B {Z^3 e^3 \over \Omega^2 m_p^2 A^2} \\
&\sim {1 - \ln{2} \over 4} {m_e^2 c^4 \over \Omega^2 m_p^2} \alpha {B \over
B_c} {Z^3 \over A^2}
\sim 3 \times 10^{-7} {Z^3 \over A^2} {B_{15} \over \Omega_{20}^2} \ll 1,
\end{split}
\end{align}
where $B_c \equiv m_e^2 c^3/e\hbar = 4.413 \times 10^{13}$ G is the quantum
critical field, $\alpha$ is the fine-structure constant and $\Omega_{20}
\equiv \Omega/(10^{20}\,\text{ergs/g})$.  The electrons
are ballistically accelerated by the electrostatic field that supports the
neutralizing ions, and Ohmic resistivity is inapplicable.
\subsection{A Possible Mechanism}
Crustal and interior motions in the neutron star rearrange its surface
magnetic field and the structure of its magnetosphere \citep{TD92,TD95}.
As a result, the current may, locally or globally, concentrate on sheets of
thickness $h \ll r$.  Concentration of currents on thin sheets occurs during
magnetic reconnection \citep{PF00}.  When contacting plates of highly
conductive neutron star crust, with frozen-in fields, are displaced with
respect to one another, volumes of magnetosphere with different fields may
be brought into contact, with the tangential magnetic discontinuity
accommodated by a thin current sheet.  In Eqs.~\ref{currentdensity} and
\ref{edensity} $r$ is replaced by $h$ which may be very small, at least near
the crustal plates.  Eq.~\ref{scatprob} becomes
\begin{align}
\begin{split}
\label{scatthin}
P_{scatt} &\sim {1 - \ln{2} \over 4} {m_e^2 c^4 \over \Omega^2 m_p^2} \alpha
{Z^3 \over A^2} {B \over B_c} {r \over h} \\ &\sim 3 \times 10^{-7}
{Z^3 \over A^2} {B_{15} \over \Omega_{20}^2} {r \over h}.
\end{split}
\end{align}

When $P_{scatt} \ll 1$ the resistivity is not Ohmic; most electron motion
is ballistic.  But when $P_{scatt} \gg 1$ the ``scattering optical depth''
of the coronal arch to the electrons is $P_{scatt}$ and the electrons
undergo a one-dimensional random walk (the only scattering possible is in
the backward direction), reducing their mean speed to $c/P_{scatt}$.  This
increases the charge density required to carry the current density by the
same factor of $P_{scatt}$, and quasi-neutrality multiplies the density of
scatterers by that factor again.  The result is a limiting current density
set by the condition $P_{scatt} \approx 1$:
\begin{align}
\begin{split}
\label{jmax}
J_{max} &\sim {\Omega^2 m_p^2 c \over \pi (1 - \ln{2}) e^3 r}{A^2 \over Z^3}
\sim 8 \times 10^{24}\ \Omega_{20}^2 {A^2 \over Z^3}\ {\mathrm{esu} \over
\text{cm$^2$-s}} \\ &\sim 3 \times 10^{15}\ \Omega_{20}^2 {A^2 \over Z^3}\
{\mathrm{A} \over \text{cm$^2$-s}}.
\end{split}
\end{align}

When the current density required by the magnetic field exceeds $J_{max}$
\begin{equation}
{c \over 4 \pi}{B \over h} > J_{max}
\end{equation}
a condition found if
\begin{equation}
{h \over r} < {1 - \ln{2} \over 4}{B e^3 \over \Omega^2 m_p^2} {Z^3 \over
A^2} \sim 3 \times 10^{-7} {B_{15} \over \Omega_{20}^2} {Z^3 \over A^2}.
\end{equation}
it becomes impossible for the current sheet to carry the current implied by
the assumed magnetic configuration.  It is difficult to be sure what the
consequences will be, but it is plausible that the result will be a runaway
increase in plasma density on the current sheet as the scattering and
resistance increase and the rapid deposition of enough magnetic energy and
the field configuration relaxes to one consistent with $J < J_{max}$.
\section{The Clumping Factor}
The extraordinary brightness of FRB requires coherent emission.  \cite{K14}
estimated the degree of clumping for emission by relativistically expanding
plasma, but in the present model the source is trapped in a neutron star
magnetosphere, with zero bulk velocity.  The angle-integrated spectral
density emitted by a charge $N_e e$ moving with Lorentz factor $\gamma \gg
1$, integrated over the passage of its radiation pattern, is \citep{J62}
\begin{equation}
\label{spectral}
{dI \over d\omega} \sim {(N_e e)^2 \gamma \over c}.
\end{equation}
Adopting $\gamma = 100$, distributing the observed $\sim 10^{40}$ ergs of
the brightest FRB \citep{T13b} over a spectral bandwidth $\sim 10^{10}$
s$^{-1}$ and $\sim 10^{10}$ radiating ``bunches'' (coherent emission is only
possible for sources whose dimensions are $< \lambda/2 \approx 10\,$cm) and
allowing for $\sim 1\,\text{ms/(r/c)} \sim 30$ passages through a current
sheet of dimension $\sim r$ yields $N_e \sim 6 \times 10^{22}$.

This would imply a potential $V \sim N_e e/(\lambda/2) \sim 3 \times
10^{12}\,$esu, or an electrostatic energy of $1.5 \times 10^3$ ergs ($10^6$
GeV) per electron ($\gamma \sim 10^9$) and an electric field of $3 \times
10^{11}\,\text{cgs}$ or $10^{14}$ V/cm.  Such an electric field would be
about 1\% of the characteristic quantum field $E_c \equiv m_e^2 c^3/e\hbar$
\citep{HE36,S51}, approaching the threshold of ``Schwinger sparks'' at 5\%
of the characteristic field \citep{SY15}.  However, these fields are
inconsistent with the assumed $\gamma \sim 100$.  

A self-consistent solution of Eq.~\ref{spectral} and $\gamma m_e c^2 =
N_e e^2 / (\lambda/2)$ is
\begin{align}
\label{eburst}
\begin{split}
N_e &= \left({dI \over d\omega}{(\lambda/2) m_e c^3 \over e^4}\right)^{1/3}
\sim 2 \times 10^{20}\\
\gamma &\sim 10^7.
\end{split}
\end{align}
The electron energy $eV \sim 10^4$ GeV, about 100 times less than in
the preceding paragraph.

These extraordinary numbers follow from the inferred brightness temperatures
(and cosmological distances) of FRB and are not specific to the model
\citep{K14}; qualitatively similar inferences follow from the observation of
nanosecond ``nanoshots'' of Galactic pulsars \citep{S04,HE07} whose sources
must be as small as a few m.  Such electrostatic potentials imply much
larger $\gamma \sim eV/(m_e c^2)$ {during outburst} than are found for
the {slowly decaying, quasi-}steady state magnetosphere in
\S\ref{eenergy}.

The plasma frequency at the electron density (\ref{edensity}) exceeds the
frequency of the observed radiation, and in a current sheet carrying the
current density (\ref{jmax}) it is orders of magnitude greater still.
Despite this, it is possible for radiation at the observed frequencies to be
emitted because the dense plasma, and radiation source, is confined on
magnetic surfaces with an abrupt density discontinuity; surface charges and
currents of this overdense plasma are the sources of radiation, which does
not propagate through the plasma itself.

Curvature radiation has been suggested as the emission mechanism of radio
pulsars \citep{G07,MY16}, with their high brightness temperature requiring,
and explained by, clumping (correlation) of the radiating electrons.
Curvature radiation may also explain their linear and circular polarization
\citep{GS90,G10,KG12,WWH14}.  The subject remains controversial.

In analogy, curvature radiation {might} account for the radio emission
of FRB.  An argument in favor of this hypothesis is the observation of both
linear \citep{M15} and circular \citep{P15} polarization in FRB, though in
different events.  {However, at the Lorentz factors estimated during
outburst (\ref{eburst}), curvature radiation would occur at frequencies far
in excess of those observed in FRB.}
%
\section{Summary and Conclusion}
Several lines of argument, including energetics, comparison of event rates
with those of other transients such as SNR and GRB, and the energy emitted
as an accreting or despinning neutron star approaches its stability limit
indicate that FRB are not produced by one-time catastrophic events.  This
conclusion is supported by the discovery \citep{T13a,C15} of a double-pulsed
FRB with subpulse separation (about 2 ms) greater than the neutron star
collapse time, indicating two distinct events produced by the same source,
and was recently confirmed by the discovery of a repeating FRB \citep{S16}.
In such models there is no connection between the observed FRB rate and the
birth rate of their parent objects.

FRB may be associated with other brief transients.  If FRB
are at the ``cosmological'' distances indicated by their dispersion
measures, giant PSR pulses (including RRAT) cannot be sufficiently energetic
unless the neutron star is both extremely fast-rotating and strongly
magnetized.  This combination is not observed among Galactic pulsars and is
almost self-contradictory, because strong magnetic fields lead to rapid
spindown.  The remaining category of known brief electromagnetic transients
is the Soft Gamma Repeaters, and in particular the sub-ms rise times of
their giant flares.  The rate of Galactic giant SGR flares exceeds the
observed rate of FRB per galaxy by about four orders of magnitude if FRB
are at ``cosmological'' distances, but there are several possible
explanations, of which the most obvious is that only a small fraction of
SGR-associated FRB are energetic enough to be observable at those distances.

Consideration of possible mechanisms for dissipation of magnetostatic energy 
in transparent regions, as required by the fast (sub-ms) time scales of both
FRB and the rising phase of giant SGR flares, points toward currents flowing
on magnetospheric arches, analogous to Solar coronal arches.  Such arches
have lifetimes of thousands of years or less, depending on the magnitude of
the magnetic field, and may explain why SGR/AXP behavior is found in young,
high-field neutron stars, and why neutron stars that are either older or
have smaller fields may be radio pulsars but are not SGR/AXP.  The criterion
is the ratio of field to age, explaining why older neutron stars do not
show SGR/AXP activity even if their fields are in the ``magnetar'' range
\citep{MHTH05}, as well as younger but lower field neutron stars like the
Crab pulsar.

I hypothesize that SGR/AXP and FRB activity require the persistence of
magnetospheric currents.  Their decay time $\tau$ (Eq.~\ref{tau}) for
neutron stars with the highest magnetic fields is consistent with estimated
SGR/AXP ages.  The smaller values of $\tau$ for neutron stars with lower
fields (such as the Crab pulsar) explain why they are not SGR/AXP, even if
they are younger than observed SGR/AXP.  Magnetospheric currents have also
decayed in older pulsars, even those with fields in the ``magnetar'' range
\citep{M03,MHTH05}, explaining why they too are not SGR/AXP.  

The fact that the event rate of FRB ($\sim 10^{-5}$/y per galaxy, if at
cosmological distances), is $\sim 10^4$ times the observed rate of giant
flares from Galactic SGR indicates that if they are associated, the
association is with a tiny minority of SGR flares, perhaps the most
energetic.  SGR 1806-20 radiated about $4 \times 10^{46}$ ergs \citep{H05},
and the rarer SGR associated with FRB at ``cosmological'' distances may be
orders of magnitude more energetic.  \cite{H05} suggested that some short
(duration $< 2$ s) GRB are actually SGR, but SGR flares are observed to have
even shorter widths of 0.1--0.2 s, excluding them as the sources of most
``short'' GRB.  The observed SGR may radiate from magnetically confined pair
plasmas \citep{K82,TD92,TD95} at photospheric temperatures of $\sim 20$ keV
\citep{K96}, while the more energetic FRB may produce relativistically
expanding pair plasmas.  Because of Doppler boosting the escaping radiation
has energies comparable to the initial MeV temperature \citep{G86},
suggesting a new class of ultra-hard, short and faint GRB coincident with
FRB.

Some alternative hypotheses and the non-detection of a FRB accompanying
SGR~1806-20 are discussed in the appendices.  \cite{BL16} recently
identified the power (\ref{power}) with the steady emission of magnetars. 
\appendix
\section{Compton Recoil of Annihilation Radiation}
In the latter ($\gtrsim 1\,$ms) part of a SGR flare its pair plasma
thermalizes, radiating the characteristic soft SGR spectra, but during their
rapid ($< 1\,$ms) rising phase more energetic pair annihilation gamma-rays
(a tiny fraction of the total energy radiated) may be produced, such as were
reported from SGR~0525-66 \citep{C80a,C80b}.  These gamma-rays Compton
scatter in surrounding plasma, creating a broadly directed flow of
semi-relativistic electrons whose distribution function peaks at the highest
kinematically permitted energy.  This excites the ``bump-on-tail'' plasma
instability in a background plasma.
\subsection{Distribution Function}
Annihilation radiation Compton scatters on non-relativistic plasma,
producing a broad beam of recoiling electrons.
In the simplest possible case the annihilation spectrum is monochromatic
with $h\nu = m_e c^2 = 511\,$keV, the target electron distribution is
``cold'', with negligible (compared to $c$) thermal velocities, the target
is sufficiently distant from the source that the photons can be considered
a directed beam and the effects of magnetic fields are ignored (valid if $B
\ll m_e c^3/e \hbar = 4.4 \times 10^{13}\,$gauss and $B$ is parallel to the
photon beam).  The differential scattering cross-section is given by the
Klein-Nishina formula \citep{BJ64}
\begin{equation}
{d \sigma \over d \Omega} = {e^4 \over 2 m_e^2 c^4}\left({\nu^\prime \over
\nu}\right)^2 \left({\nu^\prime \over \nu} + {\nu \over \nu^\prime} -
\sin^2\theta\right),
\end{equation}
where the frequencies of the incident and scattered photons are $\nu$ and
$\nu^\prime$.  Kinematics dictates a relation between the direction of the
scattered electron and its energy
\begin{equation}
\nu^\prime = {\nu \over 1 + (2h\nu/m_e c^2)\sin^2(\theta/2)}.
\end{equation}

Plasma instability depends on the component of electron velocity parallel
to the wave vector of the plasma oscillation, so consider only the component
of velocity parallel to the incident photon beam.  The resulting
distribution function of scattered electrons is shown in Fig.~\ref{compt}:
\begin{figure}
\centering
\includegraphics[width=3in]{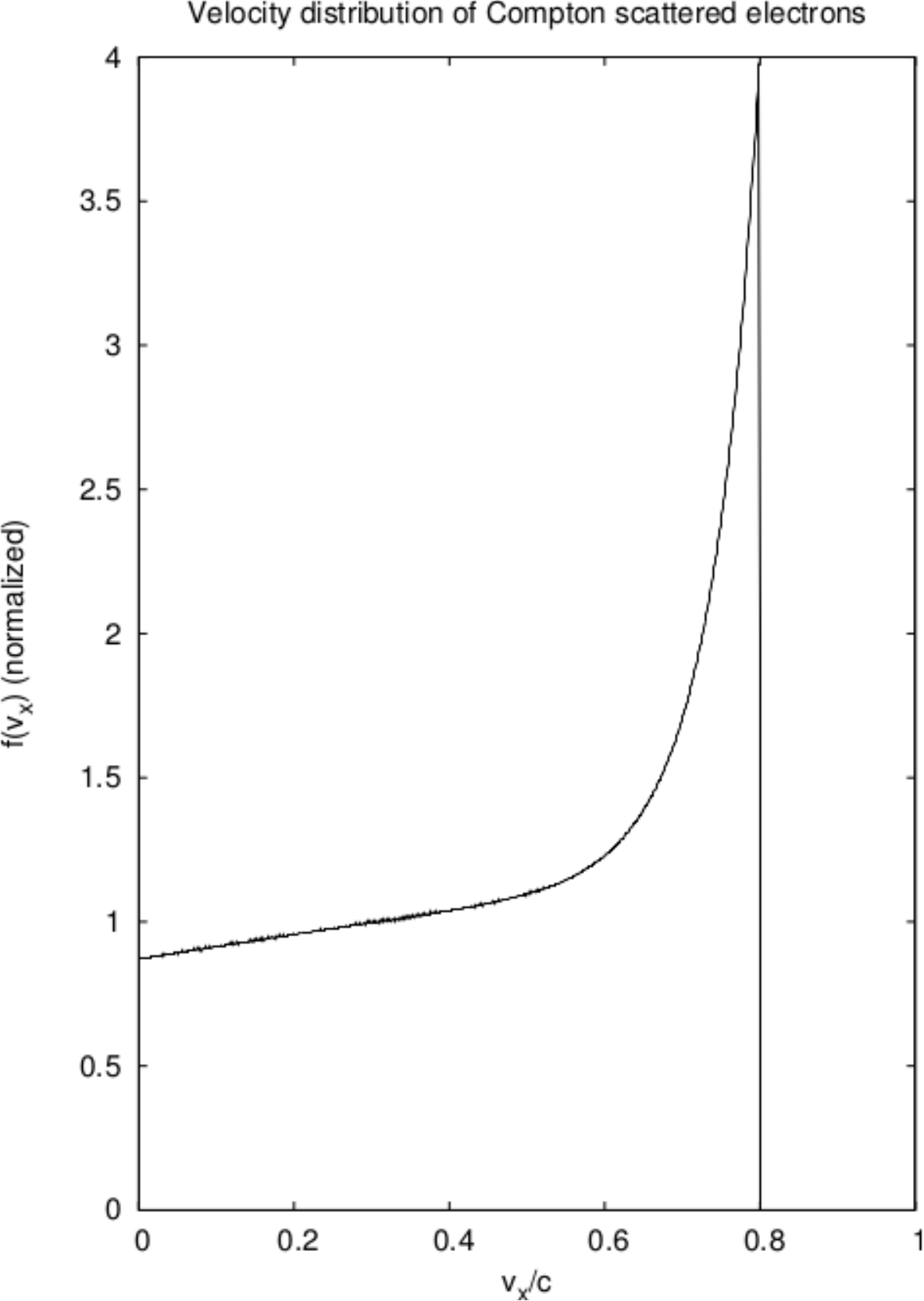}
\caption{\label{compt}Normalized distribution function {\it vs.\/} velocity
component in the beam direction of electrons Compton scattered by a beam
of 511 keV photons.}
\end{figure}

The distribution function $f(\theta,v)$ of scattered electrons evolves as a
result of Coulomb drag on the much larger population of unscattered
electrons and of ions.  This is described by a Fokker-Planck equation
\begin{equation}
{\partial f(\theta,v) \over \partial t} = S(\theta,v) - {\partial \over
\partial v} \left({dv \over dt} f(\theta,v)\right),
\end{equation}
where $S$ is the scattering source function and $dv/dt \propto \beta^{-2}
\gamma^{-3}$ represents Coulomb drag (slowing by loss of
energy to background plasma; \cite{G72}).  In steady state this reduces to:
\begin{equation}
\label{dragf}
f(\theta,v) = C \beta^2 \gamma^3 \int_{v^\prime = v}^\infty S(\theta,
v^\prime)\,dv^\prime,
\end{equation}
where $\beta \equiv v/c$ and $\gamma \equiv (1-\beta^2)^{-1/2}$.  Here $\vec
v$ is the velocity vector and $v$ the speed; $\theta$ remains constant
during slowing (aside from a small amount of straggling), so $f(\theta,v)$
may be evaluated independently for each $\theta$ and the three-space speed
projected onto the beam direction.
\subsection{Plasma Instability}
A multipeaked electron distribution function $f(v_x)$, such as that shown
in Fig.~\ref{compt} with the addition of the thermal electron peak at small
velocity (not in the Figure, that only shows the scattering source) is
unstable to the electrostatic ``bump on tail'' instability.  This
instability is essentially inverse Landau damping, and occurs whenever the
derivative of the distribution function with respect to some component of
electron velocity is positive.  Electron
plasma waves with phase velocities $v_{ph} = \omega/k$ ($\omega$ and $k$ are
the plasma wave frequency and wave vector, related by the dispersion
relation $\omega^2 = \omega_p^2 + 3 k^2 \lambda_D^2$ where $\omega_p$ is the
plasma frequency and $\lambda_D$ the Debye length) for which $\left.\partial
F_0(v_x) / \partial v_x\right \vert_{v_x = v_{ph}} > 0$ grow at a rate
\citep{C74}
\begin{equation}
\label{gr1}
\gamma = {\pi \over 2}{\omega_p^3 \over k^2} {n_{beam} \over n_e}\left.
{\partial F_0(v_x) \over \partial v_x}\right\vert_{v_x = \omega/k},
\end{equation}
where $F_0(v_x)$ is the normalized one-dimensional electron distribution
function of a density $n_{beam}$ of high velocity (Compton recoil) electrons
and $n_e$ is the background electron density.  If the fast electrons are
only a small fraction of the total electron density and their speeds are
much greater than the electron thermal velocity, conditions that are
generally met, then for waves excited by the fast electrons $\omega \approx
\omega_p$ and $k \approx \omega_p/v$, where $v$ is a velocity on the rising
part of the electron distribution function.  Then the growth rate becomes
\begin{equation}
\label{gr2}
{\pi \over 2}{\omega_p v^2 \over n_e}\left.
{\partial f(v_x) \over \partial v_x}\right\vert_{v_x = \omega/k},
\end{equation}
where $f(v_x)$ is not normalized to $n_{beam}$ but is the full electron
distribution function.

The rapid decrease of drag deceleration with increasing speed, shown by the
factor $\beta^2 \gamma^3$ in Eq.~\ref{dragf}, acts to create a minimum in
$f(v_x)$ (the large peak in $f(v_x)$ from thermal electrons at very small
velocity is not shown in Fig.~\ref{compt}) even when $S(v)$ is a
monotonically decreasing function, as it can be for power law photon spectra
or a mixture of two- and three-photon annihilation (not shown).  For
example, if $S(v) \propto v^{-n}$, making the nonrelativistic approximation
$\gamma \to 1$, then $f(v) \propto v^{3-n}$.  For all but the most steeply
decreasing $S(v)$, $n < 3$ and $f^\prime(v) > 0$ for $v$ exceeding a few
thermal velocities, and the plasma is unstable.  In the opposite limit of
scattering by very energetic gamma-rays $S(v) \to \delta(v-c)$, $f(v)
\propto v^2 \gamma^3$, $f^\prime(v) > 0$ and again the plasma is unstable
(for quantitative growth rates Eqs.~\ref{gr1}, \ref{gr2} need to be
corrected for the relativistic increase in electron mass).

This mechanism of producing unstable electron distributions may operate
anywhere an intense flux of gamma-rays, not necessarily from positron
annihilation, whose spectrum is not too steep, is incident upon a
non-relativistic plasma.  The required intensity is determined by the
competition between the growth rate $\gamma$ (Eqs.~\ref{gr1}, \ref{gr2}),
proportional to the gamma-ray flux, and the collisional damping rate of
the plasma wave that depends on the plasma density and temperature.

For a plasma frequency comparable to the frequencies
$\approx 1400\,$MHz of observed FRB, $n_{beam} \gtrsim 10^{-6}n_e$ is
sufficient to grow the instability by $\sim 10$ $e$-folds to saturation in
the sub-ms time scale of FRB.
\subsection{Energetic Failure}
This apparently attractive model fails on energetic grounds.
The total Klein-Nishina cross-section for a 511 keV photon is $2.85\times
10^{-25}$ cm$^2$, so that annihilation radiation will penetrate a plasma to
a depth of about $4 \times 10^{24}$ electrons/cm$^2$.  If electrons receive
an average energy of $m_e c^2$ the total deposited energy is about
$3 \times 10^{18}$ ergs/cm$^2$.  If more energy were deposited the electron
thermal energies would be $\gtrsim m_e c^2$, their distribution function
would not be inverted ($\partial f(v_x)/\partial v_x$ would be negative at
all $v_x$) and there would be no instability.
Over a hemisphere of a neutron star the total deposited energy must be less
than $10^{31}$ ergs, failing to account for observed FRB energies, even
assuming perfectly efficient radiation, by nine orders of magnitude.

An even stronger conclusion results from noting that the critical density
for the observed 1.4 GHz FRB radiation, above which it cannot propagate, is
$n_e = 2.4 \times 10^{10}\,$cm$^{-3}$.  A magnetosphere filled with that
density of semi-relativistic plasma would contain less than $10^{23}$ ergs
of plasma energy.
\section{Possible Alternatives to SGR?}
\label{alternatives}
A number of alternative sites of FRB have been considered, in most cases
without detailed modeling of the emission mechanisms.  Here I discuss two
such sites that, like SGR, involve neutron stars, and point out significant
unresolved questions.
\subsection{Giant pulsar pulses}
The ``nanoshots'' of a few radio pulsars are even shorter than FRB, with
durations of nanoseconds \citep{S04,HE07}, but involve energies smaller than
those of FRB by many orders of magnitude.  In fact, their instantaneous
radiated power, even during their nanosecond peaks, is less than their
pulsars' mean spin-down power.  This supports the hypothesis that, as in
classical pulsar theory \citep{G68,GJ69,G07}, they do not tap stored
magnetic energy.

Giant pulsar pulses have been suggested \citep{CSP15,CW16} as the origin of
FRB.  If their instantaneous power is less than their mean spin-down power
(no release of stored magnetostatic energy), as is always the case for
Galactic radio pulsars, then the inferred luminosity of at least one
FRB at ``cosmological'' distances ($L_{FRB} \gtrsim 10^{43}$ erg/s;
\cite{T13b}) precludes explaining FRB as giant pulsar pulses unless the
neutron stars are rotating near breakup and are very strongly magnetized and
hence very young.  Their ages must be $< E_{rot}/L_{FRB} \sim 100\,$y, where
$E_{rot} \approx 5 \times 10^{52}$ ergs is the rotational energy of a
neutron star at its rotational limit.  If intrinsic FRB pulse widths are
shorter than 1 ms (the observational upper limit) then $L_{FRB}$ is
correspondingly greater and the upper bound on the age of the most luminous
FRB correspondingly less.

The existence of such high field millisecond pulsars would not violate any
law of physics, and \cite{OG71} suggested that they make supernovae.  But
they have never been observed, may lead to a conflict between SN and FRB
rates (supernova remnant energies and the absence of pulsars in most SNR
indicate they can only be an unusual subclass of neutron star births) and
cannot explain the distribution of FRB dispersion measures \citep{K16}.
\subsection{Neutron star collapse}
The collapses of {\it accreting\/} NS cannot be
the explanation of FRB.  Most neutron stars are observed to have masses of
about 1.4 $M_\odot$, but the maximum mass of neutron stars must be at least
2.0 $M_\odot$ because a few such massive neutron stars are known.  For
accretion to push a 1.4 $M_\odot$ neutron star to collapse would require the
release of at least $\sim 0.6\,M_\odot (GM/r) \approx 10^{53}$ ergs; and one
such event per 1000 years in a galaxy implies a mean X-ray emission of about
$3 \times 10^{42}$ ergs/s.  This is about $10^3$ times the 2--10 keV X-ray
luminosity of our Galaxy \citep{GGS02}.  Only if these hypothetical
progenitors of FRB are born very close (within $\sim 10^{-3} M_\odot$) to
their stability limit can the FRB rate be reconciled with the observed
bounds on galactic X-ray emission.  This argument would not exclude the
collapse of rotationally stabilized (not accreting) NS as the result of
angular momentum loss \citep{FR14} if they, like recycled pulsars, are
inefficient (${\cal O}(10^{-3}$--$10^{-4})$ \cite{B06}) X-ray emitters.
However, their mechanism of emission of a FRB remains obscure; the time
scale of collapse (and disappearance of the NS magnetic moment) would be
$10^{-4}$--$10^{-5}$ s, corresponding to emission at tens of kHz, not the
observed frequencies $\sim 1$ GHz.

The recent discovery \citep{S16} of a repeating FRB excludes all models
based on a catastrophic event that destroys the progenitor.
\section{SGR 1806-20}
\label{1806}
\cite{TKP16} found that the Parkes telescope was observing a pulsar at the
time of the giant 27 December, 2004 outburst of SGR 1806-20, and that the
SGR was 31.5$^\circ$ above the horizon and 35.6$^\circ$ away from the beam
direction.  Analysis of the data found no evidence of a fast radio burst,
with an upper limit tens of dB lower than predicted for a Galactic FRB out
of beam \citep{K14,K16}.  This appears to contradict the suggestion that FRB
and SGR outbursts are associated, but possible explanations should be
considered:
\begin{description}
\item[Atmospheric absorption:] SGR 1806-20 deposited about 1 erg/cm$^2$ of
soft (30--100 keV) gamma-rays into the Earth's atmosphere at a depth of
roughly 5 g/cm$^2$, where the molecular density is about $2 \times 10^{17}$
cm$^{-3}$.  Approximately $2 \times 10^{10}$ electrons are produced per erg
of deposited energy, mostly distributed over a single scale height, and
their recombination or capture time is long.  Because of the high density,
their collision rate (with neutral molecules) is high, absorbing energy from
a propagating radio-frequency pulse.  A rough calculation indicates an
optical depth of about 0.02 for 1.4 GHz radiation, which is insignificant.
\item[Absorption at source:] SGR 1806-20 emitted at least $10^{46}$ ergs of
soft gamma-rays, or about $2 \times 10^{53}$ photons.  Some of these may be
absorbed in a dense surrounding cloud, and if the cloud density exceeds
$10^5$/cm$^3$ the $\sim 30$ keV photoelectrons will each produce an
additional $\sim 1000$ electrons by collision in the $\sim 1\,$s SGR
duration.  Such a cloud, with DM $\gtrsim 300$ pc/cm$^3$, could be an
effective absorber of $\sim 1$ GHz radiation, but only if it were close
enough to the SGR (within $\sim 3 \times 10^{15}$ cm) to be fully ionized.
Such a dense cloud that close to a neutron star, within a SNR, is
implausible.  In addition, only a fraction $\lesssim 10^{-2}$ of the energy
of the $\sim 100\,$ms SGR flare is emitted in its $\lesssim 1\,$s rising
edge assumed to correspond to the brief FRB.
\item[Propagation Broadening:] \cite{TKP16} consider scattering (multipath)
pulse broadening time scales in the range 14--56 ms, a range suggested by
pulsar-derived models of interstellar propagation.  However, \cite{KMNJM15}
found that the most highly dispersed pulsars have broadenings of about
$6 \times 10^4$ ms at 327 MHz, scaling ($\propto \nu^{-4.4}$) to about 100
ms at the frequency of observation of \cite{TKP16}.  Pulsar searches are
strongly biased against detection of highly broadened pulses and may be
blind to highly dispersed objects, so gamma-ray selected objects in the
Galactic plane (such as SGR) may be much more broadened than those
discovered in pulsar surveys.  Perhaps a radio burst from SGR~1806-20 was so
broadened as not to have been detectable after filtering for pulsar-like
rapid time variability.
\item[Rejection as interference:] A strong source far out of beam produces
signals of similar amplitude in each of Parke's 13 beams.  This is also a
characteristic of interference (from local sources, or entering amplifiers
by their ``back doors''), and such signals may have been rejected as likely
interference.
\item The rate of Galactic SGR is $\sim 0.1$/y, while the rate of observed
FRB {\it per galaxy\/}, assuming cosmological distances, is $\sim
10^{-5}$/y per galaxy.  Only a small subset of SGR can produce the observed
FRB, and these may differ qualitatively from SGR 1806-20.
\end{description}

\cite{TKP16} also compare the FRB rate to that of short GRB, a fraction
$f < 0.15$ of which have been suggested \citep{NGPF06,O07} to actually have
been SGR.  Because of the comparatively low sensitivity of gamma-ray
detectors, if these short GRB are associated with giant SGR outbursts at a
ratio of 1:1 then comparison of the event rates indicates that they are
detected only to a distance cutoff (assuming homogeneous distributions in
Euclidean geometry) $0.013 (f/0.15)^{1/3}$ of the FRB distance cutoff.  If
the latter is at $z = 1$ then the cutoff on SGR detection is about
$55(f/0.15)^{1/3}$ Mpc, and a nominal gamma-ray fluence sensitivity of
$10^{-8}$ erg/cm$^2$ would correspond to an (isotropic) SGR energy of
$4 \times 10^{45} (f/0.15)^{2/3}$ ergs.  This is consistent with
measurements of Galactic SGR, but uncertainty in $f$ and in the effective
detector sensitivity make this a crude comparison.

\end{document}